# The basic principles and the structure and algorithmically software of computing by hypercomplex number

Ya. Kalinovsky, Yu. Boyarinova, A. Sukalo, Ya. Hitsko

### Introduction

Hypercomplex number systems (HNS) have found wide application in various branches of science and technology [1, 2]. Theoretical and applied mechanics, navigation, cryptography, digital signal processing is far from a complete list of areas of science and technology, where the use of the methods of HNS is effective.

Operating with hypercomplex numbers, especially in symbolic form, causes considerable difficulties [3, 4], related to their multidimensionality. So, for example, to multiply two quaternions with numerical coefficients, which are four-dimensional hypercomplex numbers, it is necessary to perform 16 real multiplications and 12 additions. But the coefficients in hypercomplex numbers can be not only numbers, but also different algebraic expressions (for example, polynomials), various functions and hypercomplex numbers, including symbolic variables and coefficients. Therefore, for successful operations with such objects, specialized software is required.

The most suitable for creating this type of software are computing systems or systems of symbolic computation [5]. Moreover, many of them have the means to operate with some of the HNS. For example, with complex numbers, quaternions, Clifford algebras, etc.

Of the existing numerous computing systems, the most common are MatLab, Mathcad, Mathematica, and Maple. The authors in their work dwelled on the system of computing system Maple, as one of the most developed, accessible and easily mastered.

### The purpose of work

The aim of the work is the creation of information technology that increases the efficiency of mathematical modeling of various scientific and technical problems using hypercomplex number systems of various dimensions and scientific research in the theory of hypercomplex number systems.

### Basic principles of constructing the algorithmically software of hypercomplex computations

Since the computer algebra system Maple allows you to create private packages of various computational procedures, the algorithmically software of hypercomplex calculations (hereinafter - AS) is a package having its own identifier. AS can be called up, connected to the program and transported to other computers. From the AS procedures it is possible to generate calculation programs using the tools of the algorithmic language Maple.

AS can be installed on any computer with the Windows operating system and Maple computing system no lower than the 5th version.

The AS is open for replenishment with new procedures and editing of existing procedures by any user owning Maple.

The call and connection of the AS has the form:

**read** (" name_drive**:** \\ path\\ name AS ")

**with** (name AS)

After that, a list of AS procedures will be displayed, for example:

*[Add, AddHNS, Conjug, ConvertA, DirSum2, DirSumN, Divis, GenIso, HNSnumber, LibHNS, ListHNS, MultiDim, Norma, Rad2, Refill, RefillHNS, SearchHNS, SqrtEq, Subtr, Trans, Unit, VizHNS, VizInA, VizLibHNS, InAdd, InConvertHNS, InMulti, NameBas, NatMulti, RenamA].*

A lot of attention in the development of the AS was given to the methods and structures of data representation. As noted above, the AS is designed to operate with data in a hyper-complex form. As is known, the general form of the hypercomplex number is as follows:

$$A = a_1e_1 + a_2e_2 + ... + a_ne_n , \tag{1}$$

where $n$ - dimension HNS,
$a_i$ - algebraic expressions,
$e_i$ – elements of the basis of the HNS ("imaginary units").

We shall call this form of a hypercomplex number natural. Experience shows that operating with hypercomplex numbers in kind is rather inconvenient. This is due to the fact that different operations are performed with coefficients for the basic elements that need to be identified and identified.

Consider, for example, such a simple operation as the addition of hypercomplex numbers. Suppose, in addition to (1), that there is a number

$$B = b_1e_1 + b_2e_2 + ... + b_ne_n . \tag{2}$$

If you use the addition operation within Maple, you get the following:

$$C = A + B = a_1e_1 + a_2e_2 + ... + a_ne_n + b_1e_1 + b_2e_2 + ... + b_ne_n . \tag{3}$$

Expression (3) is not a natural form of the hypercomplex number. In it it is necessary to bring the reduction of such. In the Maple system, there is a command to cast similar ones - *collect,* but when you use it, you need to specify the variable that is used to cast similar ones. Therefore, in this case, you will have to use the *collect* command once, specifying each time the variable is used to cast similar ones:

$$C = collect((collect(collect(...(A+B),e_1),...),e_n) = (a_1+b_1)e_1 + ... + (a_n+b_n)e_n . \tag{4}$$

As can be seen, (4) is a rather cumbersome construction, especially for large dimensions of HNS. Such inconveniences associated with the use of a natural form of representation of a hypercomplex number are very many.

At the same time, the Maple system has the means to get rid of these and many other inconveniences associated with the use of a natural form of representation of the hypercomplex number. The fact is that in the natural form of the representation of a hypercomplex number, only the coefficients for the elements of the basis and their sequence number in the image of the hypercomplex number are important, that is, the hypercomplex number can be represented as a vector. However, the vector-matrix form does not fit here because the components of the matrix and the vector must be of the same type. At the same time in the Maple system there is such a form of data representation as the list - *list* - an ordered set of heterogeneous data.

To operate with data in the format of lists in Maple, there are numerous commands that allow you to specify a list, determine the length of the list, add two lists of the same length, define a list member by its serial number in the list, multiply all members of the list by any expression, etc.

The representation of a hypercomplex number in the form of a list is called a list or internal representation of a hypercomplex number. Thus, instead of (1) we use the representation

$$A = [a_1, a_2, ..., a_n] . \tag{5}$$

Then the sum of two numbers will be determined:

$$C = A + B = [a_1,...,a_n] + [b_1,...,b_n] = [a_1+b_1,...,a_n+b_n], \tag{6}$$

That is, the reduction of such symbolic coefficients in accordance with their ordinal numbers in numbers is performed by automatic internal means of Maple. Also, one command executes and multiplies all members of the list by the same expression:

$$\lambda \cdot A = [\lambda \cdot a_1, \lambda \cdot a_2, ..., \lambda \cdot a_n]. \tag{7}$$

Thus, the representation of hypercomplex numbers in the format of lists greatly simplifies the development of algorithmic-software tools. However, such a solution requires the presence in AS of procedures for mutually inverse transformation of the natural and internal forms of representation of hypercomplex numbers. Moreover, it is advisable to perform certain actions on numbers in kind. In this regard, many AS procedures provide for an output in the form of a list of two elements: the first element is the result in the list form, the second in the natural one.

So, for example, here is the work of the procedure for generating a hypercomplex number of the eighth dimension:

$> A = HNSnumber(8, a, e):$

$> A[1]$

$$[a_1, a_2, a_3, a_4, a_5, a_6, a_7, a_8]$$

$> A[2]$

$$a_1 e_1 + a_2 e_2 + a_3 e_3 + a_4 e_4 + a_5 e_5 + a_6 e_6 + a_7 e_7 + a_8 e_8$$

It also turned out to be useful to give the list format to more complex hypercomplex structures. So the Cayley table of multiplication of basic elements is represented by a three-level list structure: the top level consists of a list of table rows, the second nested level the list of table cells, the third, the lowest level, is a list of structural constants of one cell.

For example, the Cayley table for generalized quaternions in natural form has the form:

| $H_{\alpha\beta}$ | $e_1$ | $e_2$ | $e_3$ | $e_4$ |
|---|---|---|---|---|
| $e_1$ | $e_1$ | $e_2$ | $e_3$ | $e_4$ |
| $e_2$ | $e_2$ | $-\alpha e_1$ | $e_4$ | $-\alpha e_3$ |
| $e_3$ | $e_3$ | $-e_4$ | $-\beta e_1$ | $\beta e_2$ |
| $e_4$ | $e_4$ | $\alpha e_3$ | $-\beta e_2$ | $-\alpha\beta e_1$ |

,

in list form:

$[[[1,0,0,0],[0,1,0,0],[0,0,1,0],[0,0,0,1]],[[0,1,0,0],[-\alpha,0,0,0],[0,0,0,1],[0,0,-\alpha,0]],$

$[[0,0,1,0],[0,0,0,-1],[-\beta,0,0,0],[0,\beta,0,0]],[[0,0,0,1],[0,0,\alpha,0],[0,-\beta,0,0],[-\alpha\beta,0,0,0]]].$

**Structure of software**

The software of hypercomplex computations consists of the following subsystems:
- execution of algebraic operations in the HNS; manipulation of the HNSand Cayley tables;
- definitions of algebraic characteristics of hypercomplex expressions;
- storage of frequently used expressions;
- performing modular operations with hypercomplex expressions;
- visualization and service.

**The subsystem of the implementation of algebraic operations in the emergency service.**

The subsystem consists of the following procedures:

**HNSnumber(n,Name,NameBas)** - generation of a hypercomplex number of the n-th dimension;

**inAdd(A,B)** -   The addition of two hypercomplex numbers A and B in a list form;
– **Add(A,B,dimHNS)** - addition of two hypercomplex numbers A and B of dimension dimHNS in natural form;
– **Subtr(A,B,dimHNS)** - subtraction of two hypercomplex numbers A and B dimHNS dimension in natural form;
– **inMulti(A,B,HNS)** - multiplication of two numbers in the list form;
– **natMulti(A,B,HNS,nBas)** - multiplication of two hypercomplex numbers in natural form;**Divis(A,B,nameHNS)** -  division of numbers in a list form;
– **Rad2(A,Name, nameBas)** - extracting a square root of a hypercomplex number in any form;
– **SqrtEq(A,B,C,NameHNS)** - solution of a hypercomplex quadratic equation.

**The subsystem transformations of HNS and Cayley tables.**

The subsystem consists of the following procedures:
– **inConvertHNS (M,Name)** –  transformation of the Cayley table from the natural view into a list view (list of structure constants);
– **VizHNS(Spis, nam)** – visualization list of the HNS in the Cayley table with this basis;
– **nameBas(A)** – determination of  identifier of the basis of the HNS by the hypercomplex number in its natural form;
– **renamA(A,nam,dimHNS)** –  renaming of the base identifier in the hypercomplex number in a natural form;
– **VizHNS(Spis, nam)** – visualization of the HNS list in the Cayley table with this basis;
– **LibHNS()** –  storage of Cayley tables;
– **SearchHNS(nameHNS, nameRepos)** –  the procedure for searching for the HNS in the storege by its name;
– **VizInA(inA,E)** Visualization of a hypercomplex number from a list form to a natural one;
– **ConvertA(A,DimHNS)** – transformation of a hypercomplex  number from a natural form to a list;
– **Refill(Spisok, Element)** – replenishment of the list by one element;
– **ListHNS(DimHNS)** –   Generation of the template list for the internal representation of the HNS;
– **nameBas(A)** – determination of the identifier of the basis of the HNS by a hypercomplex number in its natural form;
– **renamA(A,nam,dimHNS)** –  renaming of the base identifier in the hypercomplex number in the natural form;
– **RefillHNS(nameLib, nameHNS)** –  remove HNS from the storage;
– **VizLibHNS(LibHNS)** – view all HNS in natural form;
– **Trans(M,s,t)** –  transposition of rows and columns of  the Cayley table by HNS;
– **AddHNS(Name, Table, Comment, Type)** –  the completion of HNS storage;
– **GenIzo (L,nameHNS, newBas)** –  generation of isomorphic HNS by linear transformation of the basis;
– **DirSum2(Name1,Name2)** –  the construction of a direct sum of two HNS;

- **DirSumN (Spisok, nameBas)** – the construction of a direct sum of several HNS;
- **MultiDim(nameHNS1,nameHNS2,nameBas,markKom,nameHNS)** – multiplication of the HNS dimension;
- **SysIzo(HNS1, HNS2)** – generation of a system of equations for the isomorphism of two HNS;
- **SolIzo(SysEq)** – Solution of the system of equations of isomorphism of two HNS.

**Subsystem for determining the algebraic characteristics of hypercomplex expressions**

The subsystem consists of the following procedures:
**Norma(A,nameHNS)** – the definition of the norm of a hypercomplex number in a list form;
- **Unit(nameHNS,name)** – the definition of a single element of the HNS;
- **Conjug(A,nameHNS,nam)** – construction of conjugate number;
- **Divis(A,B,nameHNS)** – division procedure;
- **HNSnumber(n,Name,NameBas)** – procedure for generating a hypercomplex number.

**The storage subsystem of frequently used expressions**

The subsystem contains ready-made formulas for performing various operations and calculations for fixed HNS. This subsystem can be replenished and saved by the user.

**The subsystem for executing modular operations with hypercomplex expressions**

The subsystem consists of procedures for constructing a system of residual classes for hypercomplex modules, determining the representability of a hypercomplex number, the Euclidean algorithm for hypercomplex numbers,etc.

**Subsystem of visualization and service.**

The subsystem consists of the following procedures:
- **VizInA(inA,E)** – visualization of a hypercomplex number from a list form to a natural;
- **ConvertA(A,DimHNS)** – converting a hypercomplex number from a natural form to a list;
- **Refill(Spisok, Element)** – replenishment of the list by one element;
- **ListHNS(DimHNS)** – Generation of the template list for the internal representation of the HNS;
- **inConvertHNS (M,Name)** – transformation of the Cayley table from a natural view to a list view (list of structure constants);
- **RefillHNS(nameLib, nameHNS)** – Procedure of remove of HNS from the storage.

Such a structure and composition of software of hypercomplex character calculations in the Maple environment, as will be shown below, makes it possible to simplify the processes of creating software for mathematical modeling of various scientific and technical problems.

**Structure of some AS procedures**

In order to show the effectiveness of the application for the representation of data of list structures, let us consider the principles of constructing some software procedures, such as procedures for converting a hypercomplex number from a natural form to a list form, the procedure for multiplying numbers in a list form, and the procedure for multiplication in a natural form.

**Principles of the procedure for converting a hypercomplex number from a natural form to a list form.**

The problem is to translate the number $A$ in the form (1) into the form (5). This can be done very simply: make substitutions $e_i = 1$, $i = 1...n$ using the *subs* command, and then convert the resulting expression $A = a_1 + a_2 + ... + a_n$ with the *convert* (A, list) command to (5). However, if in $A$ one or more of the coefficients $a_i$ are zero, that is, the hypercomplex number $A$ is incomplete, then a list is obtained, the length of which is less $n$, which will lead to incorrect results in the future.

The idea of an algorithm that works correctly with both full hypercomplex numbers and incomplete ones is as follows. First, a list of length $n$ is generated - the result of the work:

$$inA = [0,0,...,0]. \qquad (8)$$

Then a list of the same length is generated, consisting of the equalities $e_i = 0$:

$$sp = [e_1 = 0,...,e_i = 0,...,e_n = 0],$$

for each of the values of the indices a list of such equations is constructed:

$$sp1 = [e_1 = 0,...,e_i = 1,...,e_n = 0] \qquad (9)$$

and the substitution of the original number of the list (9) is performed. If in the $A$ number of components $a_i = 0$, it will be included in the list (8). Thus, this substitution allocates a coefficient $a_i$, which is assigned to the $i$-th element of the list (8).

The full text of the program is given below. The formal parameters of the procedure are: $A$ - a hypercomplex number in natural form, $DimHNS$ - the dimension of the HNS, in which the number $A$ is given.

```
ConvertA := proc(A, DimHNS)
    local inA, sp, i, sp1, e, A1;
    inA := [seq(0, i = 1..DimHNS)];
    if A <> 0 then
        A1 := subs(nameBas(A) = e, A);
        sp := [seq(e[i] = 0, i = 1..DimHNS)];
        for i to DimHNS do
            sp1 := sp; sp1[i] := e[i] = 1; inA[i] := subs(sp1, A1)
        end do
    end if;
    RETURN(inA)
end proc
```

**Principles of the procedure for multiplying hypercomplex numbers in a list form.**

This procedure multiplies two hypercomplex $A$ and $B$, which are specified in the form of lists. Multiplication is carried out in accordance with the laws of composition of the system HNS.

Thus, the procedure realizes the generalized formula for multiplying hypercomplex numbers [1, 6]:

$$A \cdot B = \sum_{i=1}^{n}\sum_{j=1}^{n}\sum_{k=1}^{n} a_i b_j \gamma_{ij}^{k}, \qquad (10)$$

where: $n$ - dimension $HNS$,

$\gamma_{ij}^{n}$ – structural constants of the laws of composition $HNS$.

The procedure calls the Cayley tables storage (procedure LibHNS ()), which looks for information about the specified HNS (SearchHNS (HNS, LibHNS ())), which contains a three-level list of the structure constants of the HNS. Then formula (10) is satisfied. The full text of the program is given below.

```
inMulti := proc(A, B, HNS)
    local X, k, i, j, Lib, HNS1;
    Lib := LibHNS( );
    HNS1 := SearchHNS (HNS, Lib);
    X := [seq(0, i = 1 ..nops(HNS1[1]))];
    for k to nops(HNS1[1]) do
        X[k] := 0;
        for i to nops(HNS1[1]) do
            for j to nops(HNS1[1]) do
                X[k] := X[k] + A[i]*B[j]*HNS1[i,j,k]
            end do
        end do
    end do;
    RETURN(X)
end proc
```

Here is an example of the call and operation of this procedure. We generate two hypercomplex numbers of the fourth dimension:

$A := HNSnumber(4, a, e)[1]; \ B := HNSnumber(4, a, e)[1];$

$A := [a_1, a_2, a_3, a_4]; \ B := [b_1, b_2, b_3, b_4];$

We construct their product in the HNS $Q4N$ - the non-commutative auto-doubling of the generalized system of complex numbers [1, 6 ± 8], whose Cayley table has the form:

| $Q4N$ | $E_1$ | $E_2$ | $E_3$ | $E_4$ |
|---|---|---|---|---|
| $E_1$ | $E_1$ | $E_2$ | $E_3$ | $E_4$ |
| $E_2$ | $E_2$ | $pE_1 + qE_2$ | $E_4$ | $pE_3 + qE_4$ |
| $E_3$ | $E_3$ | $-E_4$ | $pE_1 + qE_3$ | $-pE_2 - qE_4$ |
| $E_4$ | $E_4$ | $-pE_3 - qE_4$ | $pE_2 + qE_4$ | $-p^2 E_1 - pqE_2 - pqE_3 - q^2 E_4$ |

Call procedure:

$$C := inMulti(A, B, Q4N)$$

The result is a list of four components:

$$[a_1 b_1 + a_2 b_2 p + a_3 b_3 p - a_4 b_4 p^2, a_1 b_2 + a_2 b_1 + a_2 b_2 q$$
$$+ a_3 b_4 p - a_4 b_3 p - a_4 b_4 p q, a_1 b_3 - a_2 b_4 p + a_3 b_1$$
$$+ a_3 b_3 q + a_4 b_2 p - a_4 b_4 p q, a_1 b_4 - a_2 b_3 - a_2 b_4 q$$
$$+ a_3 b_2 + a_3 b_4 q + a_4 b_1 + a_4 b_2 q - a_4 b_3 q - a_4 b_4 q^2]$$

**The procedure for multiplying two hypercomplex numbers in natural form**

The procedure for multiplying hypercomplex numbers in natural form uses the multiplication procedure in list form. First, the number is converted from the natural form to the list form, then multiplication in the list form is performed, after which the resulting product is converted from the list form to the natural one and visualized with the given name of the basis. The text of the procedure is given below.

$natMulti := \mathbf{proc}(A, B, HNS, nBas)$
    $VizInA(inMulti(ConvertA(A, nops(SearchHNS(HNS,$
    $LibHNS())))), ConvertA(B, nops(SearchHNS(HNS,$
    $LibHNS())))), HNS), nBas)$
**end proc**

Let the numbers of the third dimension have a natural form:

$$A := a_1 e_2 + a_2 e_2 + a_3 e_3 ]; \ B := [b_1 e_1 + b_2 e_2 + b_3 e_3 ];$$

We find their product in a system of triplex numbers T[1] with a multiplication table of the form

| T | $e_1$ | $e_2$ | $e_3$ |
|---|---|---|---|
| $e_1$ | $e_1$ | $e_2$ | $e_3$ |
| $e_2$ | $e_2$ | $(e_3 - e_1)/2$ | $-e_2$ |
| $e_3$ | $e_3$ | $-e_2$ | $e_1$ |

Then their product will be
$C := natMulti(A, B, T, f)$
$$C := \left(a_1 b_1 - \frac{1}{2} a_2 b_2 + a_3 b_3\right) f_1 + \left(a_1 b_2 + a_2 b_1 - a_2 b_3 - a_3 b_2\right) f_2 + \left(a_1 b_3 + \frac{1}{2} a_2 b_2 + a_3 b_1\right) f_3$$

Here the name of the basis is changed from $e$ to $f$.
We also consider the invocation of this procedure in the HNS [1] with a multiplication table of the form

| $R \oplus C$ | $e_1$ | $e_2$ | $e_3$ |
|---|---|---|---|
| $e_1$ | $e_1$ | 0 | 0 |
| $e_2$ | 0 | $e_2$ | $e_3$ |
| $e_3$ | 0 | $e_3$ | $-e_2$ |

$C := natMulti(A, B, R \oplus C, f)$
$$C := a_1 b_1 f_1 + (a_2 b_2 - a_3 b_3) f_2 + (a_2 b_3 + a_3 b_2) f_3 \ .$$

## An example of solving a practical problem with AS

In this section, a comparison of programs will be performed to solve the vector rotation problem using the quaternion by traditional means and using AS procedures. This problem often arises in systems of orientation in space, navigation, computer animation and many others [2].

The formula for the rotation of a vector $r$ in space by means of a quaternion $q$ has the form [9]:

$$r' = qrq^{-1}, \qquad (11)$$

where:

$r, r'$ – Respectively, the initial and final coordinates of the rotated point in the quaternion form, that is, this quaternion will be a vector;

$q, q^{-1}$ - Direct and inverse quaternions, which determine the axis of rotation. All multiplications in (11) are quaternionic.

In the expression (11), the quaternion $q$ must be normalized, that is, its norm should be equal to one. If the quaternion of rotation has the form:

$$Q = a_1 e_1 + a_2 e_2 + a_3 e_3 + a_4 e_4, \qquad (12)$$

then the normalized quaternion has the form:

$$q = \frac{a_1}{\sqrt{a_1^2 + a_2^2 + a_3^2 + a_4^2}} e_1 + \frac{a_2 e_2 + a_3 e_3 + a_4 e_4}{\sqrt{a_1^2 + a_2^2 + a_3^2 + a_4^2}}. \qquad (13)$$

The geometric meaning of the elements of expression (13) is as follows:
$\theta = 2 \arccos \frac{a_1}{\sqrt{a_1^2 + a_2^2 + a_3^2 + a_4^2}}$ – angle of rotation about the axis, the direction cosines of which are equal to

$$\cos \theta_i = \frac{a_i}{\sqrt{a_2^2 + a_3^2 + a_4^2}}, \quad i = 2,3,4. \qquad (14)$$

We consider the problem of determining the coordinates of a point obtained by successive rotations of a vector about two axes: first about an axis defined by a quaternion $q$, and then near an axis defined by a quaternion $p$, as shown in Figure 1.

Рассмотрим задачу определения координат точки, полученной последовательными поворотами вектора около двух осей: сначала около оси, определяемой кватернионом, а потом - около оси, определяемой кватернионом, как это показано на рис.1.

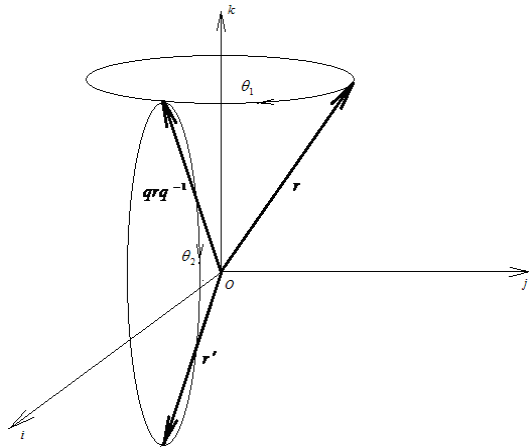

Figure 1. The rotation of a vector in space is successively about two axes

Such a complex rotation is given by

$$r' = pqrq^{-1}p^{-1}, \qquad (15)$$

Where all multiplications are quaternionic [1, 3]. The notation in (15) is as follows:
$r, r'$ – the initial and final coordinates of the rotated point in the quaternion form, that is, this quaternion will be vector;

$q, q^{-1}$ - direct and inverse quaternions defining the first axis of rotation;

$p, p^{-1}$ - direct and inverse quaternions defining the second axis of rotation.

Thus, the program should provide for the rationing of quaternions, the definition of inverse quaternions, quaternion multiplications by (15), and the simplification of the result obtained.

If the original quaternions $q$, $p$ and $r$ normalized by (13) are given, and in the list form, as well as inverse quaternions $q^{-1} = Cq$, $p^{-1} = Cp$, then one of the possible variants of the program in the form of a procedure without applying the means of the AS has the form:

```
Pov := proc(r, q, p, Cq, Cp)
    local AB, M1, M2, M3, M4, i, M, r1;
    AB := [a[1]*b[1] − a[2]*b[2] − a[3]*b[3] − a[4]*b[4], a
    [1]*b[2] + a[2]*b[1] + a[3]*b[4] − a[4]*b[3], a[1]*b[3
    ] − a[2]*b[4] + a[3]*b[1] + a[4]*b[2], a[1]*b[4] + a[2]
    *b[3] − a[3]*b[2] + a[4]*b[1]];
    M1 := subs(a[1]=p[1], a[2]=p[2], a[3]=p[3], a[4]=p[4], b
    [1]=q[1], b[2]=q[2], b[3]=q[3], b[4]=q[4], AB);
    M2 := subs(a[1]=M1[1], a[2]=M1[2], a[3]=M1[3], a[4]
    =M1[4], b[1]=r[1], b[2]=r[2], b[3]=r[3], b[4]=r[4], AB);
    M3 := subs(a[1]=M2[1], a[2]=M2[2], a[3]=M2[3], a[4]
    =M2[4], b[1]=Cq[1], b[2]=Cq[2], b[3]=Cq[3], b[4]=Cq
    [4], AB);
    M4 := subs(a[1]=M3[1], a[2]=M3[2], a[3]=M3[3], a[4]
    =M3[4], b[1]=Cp[1], b[2]=Cp[2], b[3]=Cp[3], b[4]=Cp
    [4], AB);
    for i to 4 do M[i] := factor(M4[i]) end do;
    r1 := [M[1], M[2], M[3], M[4]];
    RETURN(r1)
end proc
```

As you can see, the main volume of this program is occupied by calculating quaternion products, which are made in the form of permutations into the general formula of the product of quaternions. Since there are procedures for multiplying hypercomplex numbers in the AIC, this program is much simpler. The program using the means of the AS looks like this:

```
Pov1 := proc(r, q, p, Cq, Cp) for i from 1 to 4 do r1[i]
    := factor(inMulti(inMulti(inMulti(inMulti(p, q, H), r, H),
    Cq[1], H), Cp[1], H)[i]) :end do: RETURN(r1) end proc
```

The results of calculations for both programs are the same. So, if the problem of turning a point with coordinates $[1,2,3]$ is first solved by an angle $\theta_1 = \dfrac{\pi}{3}$, about an axis defined by $[0,1,0]$, that is, an axis $Oy$, then by an angle $\theta_2 = \dfrac{\pi}{2}$ about an axis determined by the unit vector, that is, the axis, then the final position of the point is determined by the vector quaternion

$$r' = 3e_2 + (\sqrt{3}+0.5)e_3 + (\sqrt{3}-0.5)e_4. \qquad (16)$$

**Conclusions**

The conducted researches showed that the developed software-algorithmic tools of the AS cover a large area of computations related to the modeling of processes described by hypercomplex numbers, as well as in the field of scientific research.
Their use greatly simplifies the process of software development and improves its reliability, since it uses repeatedly verified algorithms and programs. At the same time, it should be noted that the AS needs further replenishment and expansion, which is the direction of scientific research in this area.

**Bibliography**

1. Sinkov M.V., Kalinovsky Y.A., Boyarinova Y.E. Finite-dimensional hypercomplex number systems. Fundamentals of the theory. Applications. Kyiv: NAN of Ukrain, Infodruk. — 2010. — 389 p.
2. Sinkov M.V., Kalinovsky Y.O., Boyarinova Y.E . Hypercomplex numerical systems: basic theory, practical application, bibliography//NASU Data recording, storage and processing.-Preprint/-K.,-44p.

3. Kalinovsky J. A. Development of methods of theory of HNS for a mathematical modeling and computer calculations : dis. … doctor's degree: 01.05.02 / Kalinovsky J. A.; Institute for information Recording NASU. — К., 2007. — 417 p.
4. Kalinovsky Y.A. Hypercomplex numbers systems and fast algorithms for digital information processing / Y.A. Kalinovsky, D.V. Lande, J.E. Bojarinova, Y.V. Khitsko.- К:. Institute for information Recording NASU, 2014. – 130 p.
5. Von zur Gathen J. Modern Computer Algebra / J. von zur Gathen, J. Gerhard. – Cambrige: Cambrige University Press, 2013. – 808 p.
6. Kalinovsky Y.A. Study of relations between the generalized quaternions and procedure of doubling of hypercomplex numerical systems / Kalinovsky Ya.O., Boyarinova Yu.E., Sukalo A.S .- Data Rec., Storage & Processing. — 2015. — Vol. 17, № 1 – p. 36-45.
7. Kalinovsky Y. The Structure of a Hyper Fast Calculation Method Linear Convolution of Discrete Signals / Kalinovsky J.A. – 2013. – Voi. 15, # 1 –p. 31-44.
8. Kalinovsky Y.A. Some isomorphic classes for noncanonical hypercomplex number systems of dimension 2/ Y.A. Kalinovsky, D.V. Lande, Y.E. Boyarinova, Y.V. Khitsko - arXiv preprint arXiv:1403.2273, 2014.
9. Liefke H. Quaternion Calculus for Modeling Rotations in 3D Space. / [Electronic resource] / Liefke H. — Access mode: www.liefke.com/ hartmut (1998).